\documentstyle[aps12,aps,psfig]{revtex}

\begin{document}
%\draft

\title{Susceptibility Crossover Behavior in $^3$He and Xe Near Their
Liquid-Vapor Critical Point - A Progress Report\footnote{To appear in:
Proceedings 2000 NASA/JPL Investigators Workshop on Fundamental Physics
in Microgravity. D. Strayer, Editor}}

\author{ Horst Meyer}

\address {Department of Physics, Duke University, Durham, NC
27708-0305\vspace{0.1in}\\}

%\date{10/25/00}

\maketitle

\begin{abstract}

A discussion is presented on the crossover of the susceptibility from
mean-field to Ising critical behavior upon approaching  the critical
point from below and from above $T_c$, both for $^3$He and  Xe. Fits of
the experimental susceptibility data are made to curves from Monte Carlo
simulations, and the corresponding Ginzburg numbers $G_i$ for each
measured property are deduced. Also the first correction amplitudes  for
the confluent singularities are obtained from the fit of the data. The
respective ratios of these numbers and those obtained for the coexistence
curves for $^3$He and Xe, presented elsewhere, are discussed in terms of
predictions.

\end{abstract}

\section{Introduction}

The  interest in crossover phenomena from the asymptotic to mean-field
critical behavior in fluids  has been described in detail in a recent
review article by Anisimov and Sengers
\cite{Anisimov:S:00}, which lists many references and where different
theoretical approaches  and also a comparison with some experiments are
presented. The subject of a recent paper\cite{Luijten:M:00} was a
comparison between predictions for the crossover  from Monte Carlo
calculations\cite{Luijten:B:98} and experimental data of simple fluids.
The susceptibility $\chi^+$ above $T_c$ (or compressibility) and the
liquid and vapor densities along the coexistence curve (CXC) for Xe and
$^3$He were studied. From a fit of the data to the predicted
curves, the corresponding Ginzburg numbers G could be estimated. For the
CXC, the exit of the fluid from the critical regime into a background
behavior could be clearly seen by a systematic  departure from the
predicted curve, well before the regime of mean-field critical behavior
could be reached. For $\chi^+$, the behavior of Xe, a ``classical fluid",
agreed well with predictions, but there were systematic differences for
$^3$He, and a qualitative discussion was made in terms of the interplay
between quantum and critical fluctuations for this fluid.

The purpose of this progress report is to extend the same analysis to the
susceptibility $\chi^-$ data below $T_c$ for Xe and $^3$He, and also to
give a status report on this program. After a background review, the
susceptibility data of several experimental groups, namely along the
critical isochore ($T>T_c$) and along the liquid and the vapor side of
the CXC ($T<T_c$), are discussed. Comparison is made with curves from
Monte Carlo calculations
\cite{Luijten:B:98}, and the corresponding Ginzburg numbers G($\chi^-$)
are estimated. The internal consistency for the Ginzburg numbers so
obtained is checked by determining
$G$ from the fit of data to a 2-term  series expansion representing part
of the curve obtained from the  MC calculations. From the collection of
Ginzburg numbers obtained so far [$G(\chi^+)$,$G(\chi^-)$  and
$G(CXC)$],   their ratios are discussed in the light of predictions. In
spite of the uncertainties in the G($\chi^-$) below
$T_c$ due sparsity of data and experimental scatter, and also in the
G($\chi^+$) for $^3$He,   some preliminary conclusions can be made. This
progress report is to draw attention to the interest of such results, to
their  present incomplete understanding and to the great need of better
data.

\section{A short review}
\subsection{Properties considered}

We now list the properties discussed in this paper, and introduce the
definitions of reduced temperature and density,  $t \equiv (T-T_c)/T_c$
and $\Delta \rho \equiv (\rho - \rho_c)/\rho_c$. The coexistence curve is
expressed by
%-----------------------------------------------------------
\begin{equation}
\label{coex}%-----------------------
\Delta \rho_{LV} =(\rho_{\rm liq}- \rho_{\rm vap})/\rho_c = B_0
(-t)^{\beta}[1 + B_1(-t)^{\Delta_1} +B_2(-t)^{\Delta_2} ...]
\end{equation}%---------------------------------------------
%----------------------------------------------------------- 
where $\rho_{\rm liq} , \rho_{\rm vap}$ and $\rho_c $ are the densities in the
coexisting liquid and vapor phases, and at the critical point. Furthermore
$\beta$ = 0.326 is the critical exponent and the bracket includes the
correction-to-scaling confluent singularity terms. Here the $B_i$'s are
amplitudes  characteristic of the fluid, and $\Delta_1$ = 0.52, $\Delta_2$
= 1.04 are the exponents obtained by Wegner \cite{Wegner:72} and by
Newman and Riedel\cite{Newman:R:84}.

The susceptibility $\chi$ of the fluid, namely  the analog of the
susceptibility of a magnet, is given by $\chi \equiv (\partial
\rho/\partial
\mu)_T =
\rho^2\beta_T$, where $\beta_T$ is the isothermal compressibility and
$\mu$ is the chemical potential. As discussed by Sengers and Levelt
Sengers\cite{Sengers:LS:78}, the 3-D lattice-gas model (which corresponds
to the 3-D Ising model in magnets) has properties that adequately describe
real fluids. One particular aspect is that of symmetry in the $\mu-\Delta
\rho$ plane. (In this respect, $^3$He is the fluid that best conforms to
this model. See Appendix) As a consequence, the derivative
$\chi$ is a symmetric function of $\Delta  \rho$ along an isotherm. Hence
below
$T_c$, one obtains $\chi_{Liq} = \chi_{Vap}$, where the susceptibilities
are measured on both sides of the coexistence curve, and therefore we
expect consistency between the data on both liquid and the vapor sides.
Here we introduce the reduced quantity $\chi^* \equiv
\chi(P_c/\rho_c^2)$, where the critical parameters have been listed in
ref.\cite{Luijten:M:00}. Above
$T_c$ and along the critical isochore,  $\chi^* = \beta_T P_c$.

Similarly to Eq.1, the expansion for the susceptibility
$\chi^{*(+,-)}$ from the asymptotic critical regime is given by
%-----------------------------------------------------------
\begin{equation}
\label{Wegner}%-----------------------
 \chi^{*(+,-)} = \Gamma^{(+,-)}_0|t|^{-\gamma}[1 + \Gamma^{(+,-)}_1
|t|^{\Delta_1} + \Gamma^{(+,-)}_2 |t|^{\Delta_2} +.......]
\end{equation}%---------------------------------------------
%----------------------------------------------------------- 
where the indices + and - indicate the region $t>0$ along the critical isochore
and  $t<0$ along the coexistence curve, respectively. Here again
 the $\Gamma_i$'s are amplitudes  characteristic of the fluid and
$\gamma$ = 1.24 is the critical exponent. The ratio
$\Gamma_0^+/\Gamma_0^-$ for a given fluid has been  calculated from
series expansion by Liu and Fisher to be\cite{Liu:F:89}
%-----------------------------------------------------------
\begin{equation}
\label{liu/Fisher}%-----------------------
 \Gamma^+_0 / \Gamma^-_0 = 4.95 \pm 0.15
\end{equation}%---------------------------------------------
%----------------------------------------------------------- 
This compares with the value of 4.82 obtained from the ratio
$(\gamma/\beta)[(1-2\beta)\gamma/2\beta(\gamma-1)]^{\gamma-1}$ predicted
from the parametric representation of the equation of
state\cite{Schofield:L:H:69}, where
$\gamma$ = 1.24 and $\beta$ = 0.327 were used. The most recent values of
this ratio (see \cite{Bervillier}) are very close to 4.77. Predictions for
the ratio of the amplitudes
$B_1$,
$\Gamma^+_1$ and
$\Gamma^-_1$  will  be presented below.

\subsection{Ginzburg numbers and amplitude ratios}

The Ginzburg criterion and Ginzburg number have been  discussed in detail
in the article by Anisimov, Kiselev, Sengers and
Tang\cite{Anisimov:K:S:T:92} on the crossover approach to global critical
phenomena in fluids. The Ginzburg number $G$ is seen as a dimensionless
temperature, obtained from the criterion
$t \gg G$  which gives an estimate for the range of
$t$, where the classical critical theory is valid, this is where the
fluctuation contribution is small. For fluids, an order of magnitude
estimate\cite{Anisimov:K:S:T:92} of $G$ leads to $\approx 10^{-2}$, and
furthermore for a 3-D fluid, one finds $G \propto R^{-6}$, where $R$ is
the normalized molecular interaction range
\cite{Anisimov:K:S:T:92}. Hence the asymptotic critical behavior takes
place for $t \ll G$ while the classical critical behavior is expected for
$1 \gg t
\gg G$. However as pointed out in ref.\cite{Anisimov:S:00}, in ordinary
fluids the crossover is never completed in the critical domain ($t
\ll 1$) since $R$ is of the same order as the distance between molecules.
The Monte Carlo algorithm developed by Luijten and
Bloete\cite{Luijten:B:95}  allows the full crossover region in 3-D
Ising models to be covered. The calculation then gives a  curve
of a given singular property $f_i$ = $f_i( |t|/G_i)$ covering $\approx$ 8
or more decades in $|t|/G_i$, and is clearly more complete than the
expansion series expressed in correction-to scaling terms of series. A
simple check for the internal consistency in determining G can be made by
the expected expansion in terms of the corrections-to-scaling confluent
singularities as
%-----------------------------------------------------------
\begin{equation}
\label{Wegner}%-----------------------
 f_i = A_{0,i}|t|^{-\lambda_i}[1 + A_{1,i}(|t|/G_i)^{\Delta_1}  +.......]
\end{equation}%---------------------------------------------
%----------------------------------------------------------- 
where the amplitude $A_{0,i}$ is non-universal but where the numerical coefficients $A_{1,i}$, and the exponent
$\lambda_i$ are  universal for all fluids and characteristic of the
property (susceptibility, CXC etc...). A fit of the curves  $f_i$ = $f_i(
|t|/G_i)$, calculated by the Monte Carlo approach\cite{Luijten:B:98}, to
Eq. 4 (restricted to the region of $|t|/G_i$ where higher terms are
negligible) gives
$A_1(\chi^+)$ = 0.10,
$A_1(\chi^-)$ = 0.65, and $A_1(CXC)$ = 0.23.

Of particular interest here is the calculation of the susceptibility
$\chi$, both below and above
$T_c$, represented in a very sensitive way by the plot \cite{Luijten:B:98}
of the effective exponent
$\gamma_{eff}$, which is given by the derivative
%-----------------------------------------------------------
\begin{equation}
\label{differential}%-----------------------
\gamma_{\rm eff} \equiv -{{d\ln \chi_T^* }
\over {d\ln |t|}}.
\end{equation}%---------------------------------------------
%----------------------------------------------------------- 
In Fig.1, the plots of
$\gamma^{(+,-)}_{eff}$ versus
$ |t|/G^{(+,-)}(\chi)$ for $t>0$ and $t<0$, as obtained from Monte Carlo
calculations, are presented side-by-side for comparison. The various
symbols denote the successively larger values of the interaction range
$R$ that were used to generate the master curve as the distance from
$T_c$ is increased. The lines labeled ``BK, BB and SF" for
$t>0$ and ``App" for $t<0$ are theoretical curves described in
ref.\cite{Luijten:B:98}. The slope of the exponent,$-\partial\gamma_{\rm
eff}/\partial\ln|t|$, gives information on the crossover width between
the asymptotic Ising value of
$\gamma = 1.24$ and the mean-field one $\gamma = 1$. This width is shown
to be much narrower for the region
$t<0$ than for $t>0$, as was  pointed out in ref.\cite{Luijten:B:98}. We
shall see that this difference in crossover width is reflected in the data
for $^3$He.

\begin{figure}[t]
\center{\parbox{7in}{\psfig{file=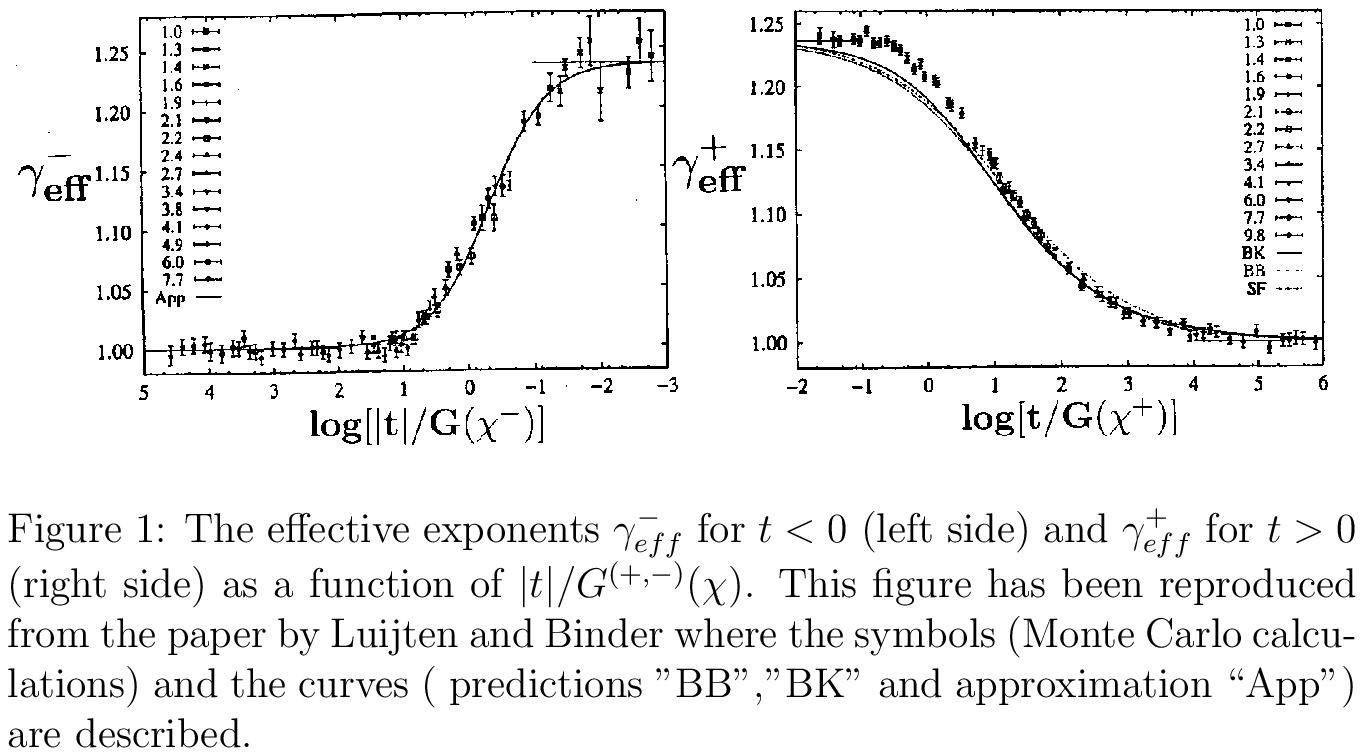,width=7in}}}
\label{fig:fig1}
\end{figure}

The relations between the G's and the first Wegner terms in the correction
to scaling  follow from Eqs, 1, 2 and 4. For instance in the case of the
susceptibility above $T_c$, one has
%-----------------------------------------------------------
\begin{equation}
\label{ginzburgWegner}%-----------------------
\Gamma_1^+ =  A_1 (\chi^+)[G(\chi^+)]^{-\Delta_1}
\end{equation}%---------------------------------------------
%----------------------------------------------------------- 
with $\Delta_1$ = 0.5, which will be used in the discussion of the data
analysis.

\begin{figure}[h]
\center{\parbox{6.2in}{\psfig{file=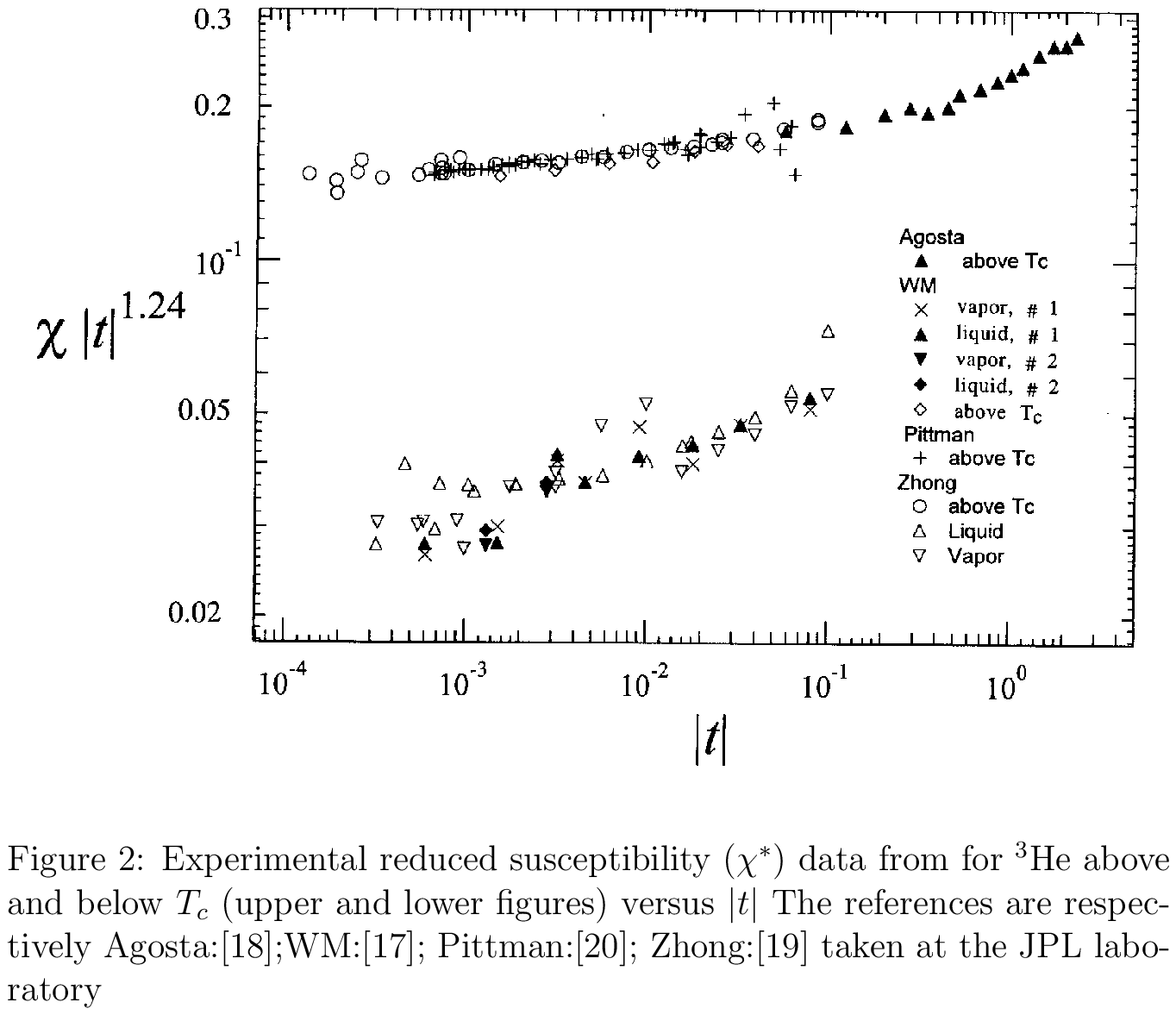,width=6.2in}}}
\label{fig:fig2}
\end{figure}

 Aharony and Ahlers\cite{Aharoni:A:80} have discussed the ratios of the
amplitudes of the ``correction-to-scaling confluent singularity" terms in
expressions such as Eqs. 1 and 2 for different properties, and in
particular for the order parameter (here CXC)  and the susceptibility
above
$T_c$ versus
$|t|$. They expressed thermodynamic  quantities with singularities at the
critical point as
%-----------------------------------------------------------
\begin{equation}
\label{Aharony}%-----------------------
 f_i =  A_{0,i}|t|^{-\lambda_i}[1
+ a_i |t|^{\Delta_1} + O|t|^{2\Delta_1}]
\end{equation}%---------------------------------------------
%-----------------------------------------------------------
 where  $\lambda_i$ is the (asymptotic) critical exponent of the property
$i$, such as susceptibility, specific heat, order parameter etc.. and the
$a_i$'s are the amplitudes of the first correction-to-scaling term of the
confluent singularity (already introduced here as
 $\Gamma_1$ and $B_1$).   Among the relations they derived, one  which
considers the ratio of the correction term amplitudes for two properties
$i$ and $j$ of a fluid is of particular interest to us, namely
%-----------------------------------------------------------
\begin{equation}
\label{effectiveexponent}%-----------------------
(\lambda_{i, eff} - \lambda_i)/(\lambda_{j, eff} - \lambda_j) = a_i/a_j
\end{equation}%---------------------------------------------
%----------------------------------------------------------- 
Here $\lambda_{i,{\rm eff}}$ is the effective exponent and
$\lambda_{i\, {\rm or}\, j}$ is the asymptotic exponent with
$\lambda_i = \gamma = 1.24$ and $-\lambda_j = \beta = 0.326$.  The
numerical values for
$\lambda_{i,{\rm eff}}$ are obtained by fitting experimental data to a
simple power law over the same range of $|t|$ where the fit to Eq. 7 has
been made. A prediction of the ratio of Ginzburg numbers via Eqs.4 and 7
can therefore be made and compared with that from experiments. One has
%-----------------------------------------------------------
\begin{equation}
\label{Ginzburgratios}%-----------------------
 ( G_j/G_i) = [a_iA_1(j)/a_jA_1(i)]^{1/\Delta_1}.
\end{equation}%---------------------------------------------
%----------------------------------------------------------- 
Here $A_1(j)/A_1(i)$ is the ratio of the numerical coefficients in Eq.4 for
the properties $j$ and $i$, listed after Eq.4.

Bagnuls, Bervillier, Meiron and Nickel\cite{Bagnuls:B:M:N:87} have
calculated the ratios $a_i/a_j$ using ``massive field theory" for the
$\Phi^4$ model in 3-D for the n=1 class. These ratios are universal and
are found to be
$a(\chi^+)/a(\chi^-)$ =
$\Gamma_1^+/\Gamma^-_1$ = 0.315 $\pm$ 0.013 and $a(\chi^+)/a(CXC)$ =
$\Gamma^+/B_1$ = 0.9$\pm$ 0.2. (See Tables VIII and IX of
ref\cite{Bagnuls:B:M:N:87}). This implies that the ratio of the Ginzburg
numbers is universal too. From Eq.9 one then obtains
G($\chi^+$)/G($\chi^-$) = 0.23$\pm$0.01 and G($\chi^+$)/G(CXC) = 0.15
$\pm$0.07.

\section{Experimental data and determination of $T_c$}

In the experiments, $\chi$ has been determined either from the intensity
of light scattering (Xe)\cite{Guettinger:C:81,Smith:G:B:71} or from the
measurements of the density versus pressure along isotherms in
Xe\cite{Michels:W:L:54}, and in
$^3$He\cite{Wallace:M:70,Chase:Z:76,Agosta:W:C:M:87,Hahn:B:Z:00} and from
the vertical density gradient in the gravity field for
$^3$He\cite{Pittman:D:M:79} above $T_c$. Here we briefly comment on the
various measurements and their respective scatter.

 As will be seen below, the $\chi$ data for $t>0$ have an appreciably
higher accuracy than those below
$T_c$. In the light scattering measurements, this might possibly be due to
the added difficulty of  sending the laser beam alternatively into  the
superposed liquid and vapor phases of a  cell with a small height, and
where the meniscus will become  concave as $T_c$ is approached, because
of the decreasing surface tension. This contrasts with measurements above
$T_c$ where the beam is positioned near mid-height  of the cell, and
where the maximum light scattering intensity is observed by slowly
scanning the vertical position. In the present analysis, the data by Smith
et al\cite{Smith:G:B:71} above
$T_c$ have not been used, because their scatter is larger than that of the
more recent data by G\"uttinger and Cannell\cite{Guettinger:C:81}. However
we note that in ref\cite{Smith:G:B:71} the amplitude ratio is
$\Gamma^+_0/\Gamma^-_0$ = 4.1, which is not far from the predictions. If
the data of ref\cite{Guettinger:C:81} above $T_c$ are combined with those
below
$T_c$\cite{Smith:G:B:71}, a ratio of 5.8$\pm$ 0.4 is obtained by fitting
both sets of data to Eq. 2. The amplitudes are listed in Table I.

The maximum value of $\chi^+$ at $\rho_c$ from isotherm data above $T_c$
is obtained with a higher precision than is the extrapolation of $\chi^-$
to the coexistence curve below $T_c$. As mentioned above, it is expected
from the Ising model that $\chi_{vap} = \chi_{liq}$. Yet there can be
appreciable scatter in both determinations which  reflects the
uncertainties both in the differentiations of the $\rho(P)$ data sets
and also in the factor $\rho^2_{Vap}$, or
$\rho^2_{Liq}$. The results are presented as reduced quantities $\chi*
\equiv \chi P_c/\rho_c^2$, where the critical parameters have been listed
in ref.\cite{Luijten:M:00}

In the experiments with Xe where optical methods were used, the
determination of the critical temperature has been achieved for the
$\chi$ measurements by observing the disappearance of the meniscus, and
by a fit to Eq.2 \cite{Guettinger:C:81}. The coexistence curve data were
fitted to Eq.1 \cite{Naerger:B:90}. On the average these determinations
were made with an uncertainty of  $\delta t\approx \pm 5
\times 10^{-6}$.

By contrast in the experiments with $^3$He without optical access, the
uncertainty  in $T_c$ is more important. It is probably smallest in the
experiments by Pittman \underline{et} \underline{al.}
\cite{Pittman:D:M:79} where $T_c$ was determined principally from
measurements below $T_c$ (coexistence curve), and where $T_c$ was obtained
from a fit of Eq.1, as described in that paper. Here the claimed
uncertainty is  $\delta T_c/T_c \approx \pm 9\times 10^{-6}$. In the
older measurements of $\chi$ from isotherms by Wallace and
Meyer\cite{Wallace:M:70}, the choice of $T_c$ was obtained by
extrapolation of both coexistence curve data and compressibility data,
and based on a simple power law with effective exponents. The uncertainty
was claimed to be
$\delta T_c/T_c \approx
\pm 6 \times 10^{-5}$. The  use of this power law led to a systematic
error in $T_c$ which was evidenced by deviations from the compressibility
data of ref.\cite{Pittman:D:M:79}, as shown in Fig.1 of
\cite{Pittman:D:M:79}. In the
$\chi$ measurements  by Chase and Zimmerman\cite{Chase:Z:76}, also from
isotherms, the determination of
$T_c$ was done in a similar way as in ref\cite{Wallace:M:70}. In the most
recent measurements of $\chi$ from isotherms by the MISTE team at
JPL\cite{Hahn:B:Z:00}, the value of $T_c$ was determined from a fit of the
$\chi$ data above $T_c$ to Eq. 6 with a systematic uncertainty of $\delta
T_c/T_c
\approx
\pm 1.5 \times10^{-5}$ in $T_c$ (F. Zhong, private communication).

\section{Data presentation and Monte Carlo calculations}

In Fig.2 the  susceptibility data for $^3$He from refs
\cite{Wallace:M:70,Pittman:D:M:79,Hahn:B:Z:00,Agosta:W:C:M:87} are
presented, both above and below $T_c$, scaled by the leading singularity
$|t|^{-\gamma}$. The data of ref\cite{Chase:Z:76} lie systematically up to
20\% below the other data sets and have not been included in the plot to
avoid overcrowding the figure. The data of ref\cite{Pittman:D:M:79} can be
made to agree well with those of ref.\cite{Hahn:B:Z:00}, if their
respective values of
$T_c$ are slightly shifted well within the stated uncertainty mentioned
above. The shifts are as follows: $\delta T_c/T_c = + 6 \times 10^{-6}$
for data of ref.\cite{Hahn:B:Z:00} and $\delta T_c/T_c = - 6 \times
10^{-6}$ for data of ref.\cite{Pittman:D:M:79}. By combining the two sets
of data with the mutually shifted $T_c$, a fit to Eq. 2 gives
$\Gamma^+_0$ = 0.145.
 For a given set of experiments,where data above and below $T_c$ were
obtained, the same choice of
$T_c$  was  implemented. The error in $T_c$ in the experiments of
ref\cite{Wallace:M:70} was corrected by an appropriate shift $\delta T_c$
withing the stated uncertainty, which resulted in the data above $T_c$ to
lie uniformly $\approx$ 5\% below those of refs.\cite{Pittman:D:M:79}  and
\cite{Hahn:B:Z:00}.

Table I lists the  amplitudes of the leading terms  $\Gamma^+_0$,
$\Gamma^-_0$ and
$B_0$, and of the first correction terms  $\Gamma^+_1$, $\Gamma^-_1$ and
$B_1$  obtained by a fit of the data to Eq.1 resp. Eq. 2. The errors
listed are all systematic, not statistical. The corresponding sources of
data are listed in the last column. The data fits for $\Gamma^+_1$ and
$\Gamma^-_1$ in
$^3$He and for
$\Gamma^-_1$ in Xe (with very scant data and appreciable scatter close to
$T_c$) were made by setting the higher terms in Eq. 1 to zero. To this
purpose, the fitting was restricted to the range
$|t| < 2\times 10^{-2}$, where presumably the higher terms in Eqs. 1, 2
(and therefore also in Eq. 4) can be neglected.  Because of the strong
correlation between the amplitudes in the data fitting procedure to Eqs 1
and 2,  an uncertainty of only say O($\pm $5\%) in
$\Gamma_0$ can produce a much larger one  of O($\pm$50\%) in
$\Gamma_1$. For the CXC of
$^3$He and Xe, and for $\Gamma^+_1$ of Xe, the amplitudes listed in
refs\cite{Guettinger:C:81,Pittman:D:M:79,Naerger:B:90} were used.

The $^3$He data by Pittman \underline{et} \underline{al.}, which result
from measuring the density difference between two superposed sensors,
show a smaller  scatter close to
$T_c$ than those ref.\cite{Hahn:B:Z:00}, but they are restricted to the
range
$t> 5\times 10^{-4}$, below which the vertical density profile becomes
sharply non-linear  as stratification from gravity increases. Above
$t> 5\times 10^{-2}$, where $\chi$ has become small, this method is no
longer sensitive, as shown by the rapidly increasing scatter.  One notes
that
 for the $^3$He $\chi^-$ data, the leading amplitude
 $\Gamma^-_0$ is consistent well within the large scatter with  the
expected
$\Gamma^-_0 = \Gamma^+_0$/4.95 = 0.029 where the factor 4.95 was given in
Eq. 3. Here we have taken $\Gamma^+_0$ = 0.145, the value from the
combined set of data from refs. \cite{Pittman:D:M:79} and
\cite{Hahn:B:Z:00}.  In spite of the data scatter below $T_c$, the
difference in the change of
$\chi\,|t|^{1.24}$   with $|t|$ in the regime above and below
$T_c$ is quite striking :
$\mathcal{R}$(t=0.1)/$\mathcal{R}$(t$\rightarrow 0)$ = 1.25 compared to
$\mathcal{R}$(-t=0.1)/$\mathcal{R}$(t$\rightarrow 0)$ = 1.93 , where
$\mathcal{R}$(t) $\equiv
\chi|t|^{1.24}$. This is consistent with the finding that
$\gamma_{eff}(t<0)$ is smaller than
$\gamma_{eff}(t>0)$ over the common experimental range
$10^{-3}<|t|< 10^{-1}$. As was mentioned before, MC calculations predict
that the crossover width is narrower for
$t<0$ than for $t>0$.

In Figs. 3 and 4,  plots of the scaled reduced and normalized
susceptibility $ \bar \chi\,
 |t|^{1.24}$ versus $|t|/G(\chi^{(+,-)})$ are shown. Here
$\bar \chi = \chi^* / \Gamma^{(+,-)}_0$,  as in ref.\cite{Luijten:M:00}.
In each figure the subscripts indicate the phase  (vapor or liquid) along
the coexistence curve.  In the limit
$|t| \rightarrow 0$ the ratio $\bar\chi\, |t|^{\gamma}$ becomes unity.
However the MC calculations for $t<0$ have more scatter than for $t>0$ as
seen from Fig.1. In spite of
this, their trajectory for $[(-t/G] < 1\times 10^{-2}$ can be estimated
quite well, since with decreasing
$|t|$ the curves will follow Eq.4, and  tend to unity.  The  values of
the $\Gamma$'s and
$G(\chi^+)$ obtained via Eq. 6 and similarly of
$G(\chi^-)$ and $G(CXC)$ for both fluids are shown in Table 1.

In Fig.3  both the
susceptibilities\cite{Guettinger:C:81,Michels:W:L:54,Smith:G:B:71} above
and below
$T_c$ for Xe are presented, - the first one already shown in
ref\cite{Luijten:M:00}. Below
$T_c$,  there are few data points, and the scatter prevents a precise
determination of
$\Gamma^-_0$ and therefore the resulting value of G($\chi^-)$ is much
more uncertain than that of G$(\chi^+)$.  It should be mentioned, as was
done in ref.\cite{Luijten:M:00}, that the fit for $t>0$ was made taking
$\Gamma_0$ = 0.0594 instead of 0.0577 obtained in
ref.\cite{Guettinger:C:81}.  This choice of
$\Gamma_0$, determined by the adopted value of $\gamma$ = 1.240, is no
doubt responsible for the different values of G($\chi^+$) obtained from
the data fit to Eq.2 and  to the MC curve, respectively 0.006 amd 0.018,
as reported in ref\cite{Luijten:M:00}.

Fig.4 shows the plots for $^3$He of refs
\cite{Pittman:D:M:79,Hahn:B:Z:00,Wallace:M:70,Agosta:W:C:M:87} of Fig.2,
with
$\Gamma^+_0$ = 0.145. For
$\chi^-$,  the value $\Gamma^-_0$ = 0.029 mentioned above was used, which
``anchors" the data presentation
$\bar \chi\,t^{1.24}$ in the asymptotic regime. The fit according to Eq.2
is restricted to $|t|< 3 \times 10^{-2}$. The top plot is different from
that of ref\cite{Luijten:M:00} as it combines the data of
refs\cite{Hahn:B:Z:00} and \cite{Pittman:D:M:79} as has been described
above. The new value of $\Gamma^+_0$ = 0.145 (instead of 0.139) then
leads to a larger value of G($\chi^+$), listed in Table 2. [By accident,
in the bottom plot for
$t<0$, the symbols WM(vapor) and JPL (liquid) on one hand, and WM(liquid)
and JPL(vapor) on the other, are undistinguishable]. It is clear that the
restricted data range in $|t|/$G$(\chi^-)$ for
$^3$He below $T_c$ does not enable  confirming the variation of the
quantum fluctuations in the crossover region, proposed for
$t>0$\cite{Luijten:M:00}, where the data extend beyond $t=10^{-1}$.

\begin{figure}[htb]
\center{\parbox{7in}{\psfig{file=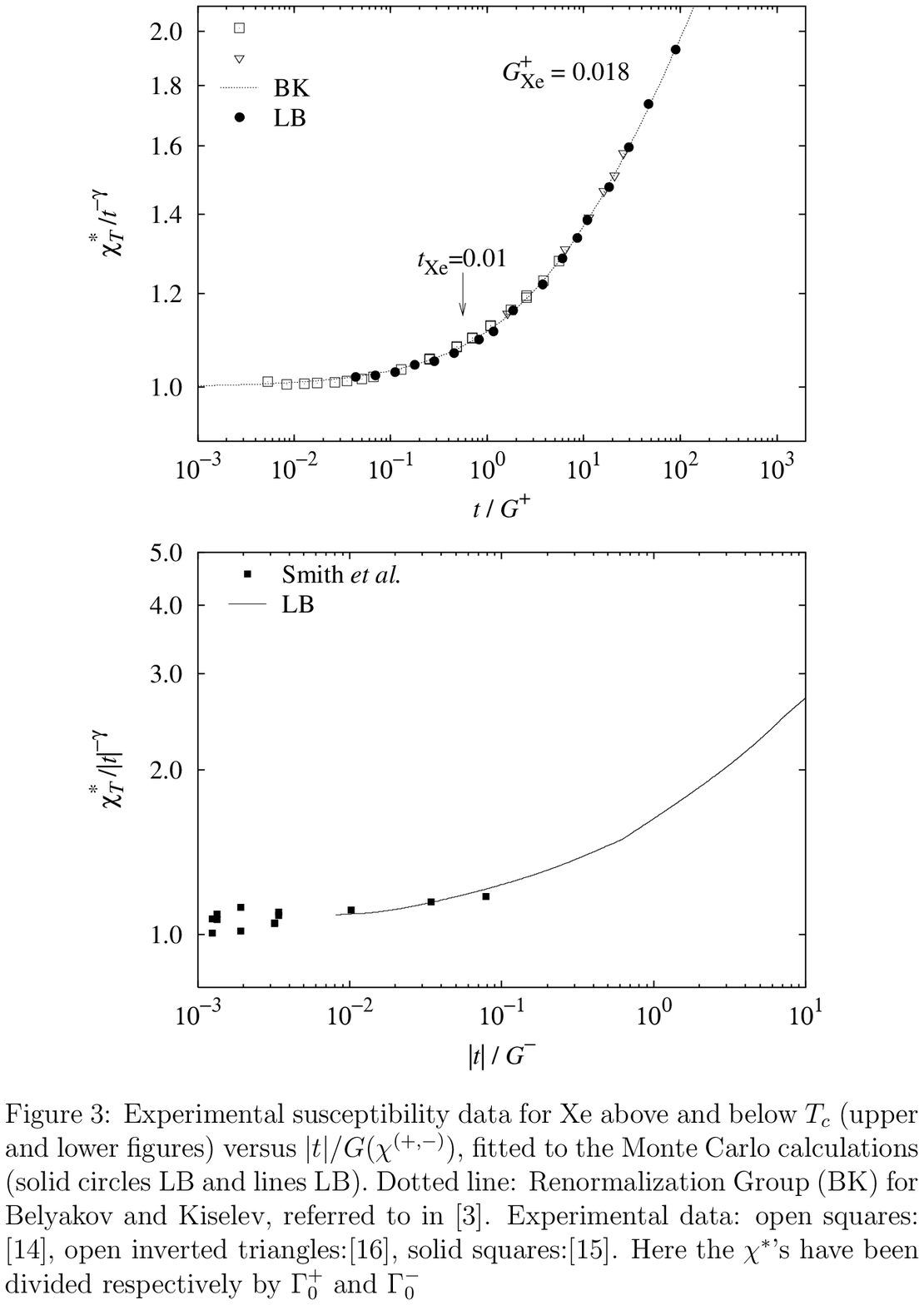,width=7in}}}
\label{fig:fig3}
\end{figure}

\begin{figure}[htb]
\center{\parbox{7in}{\psfig{file=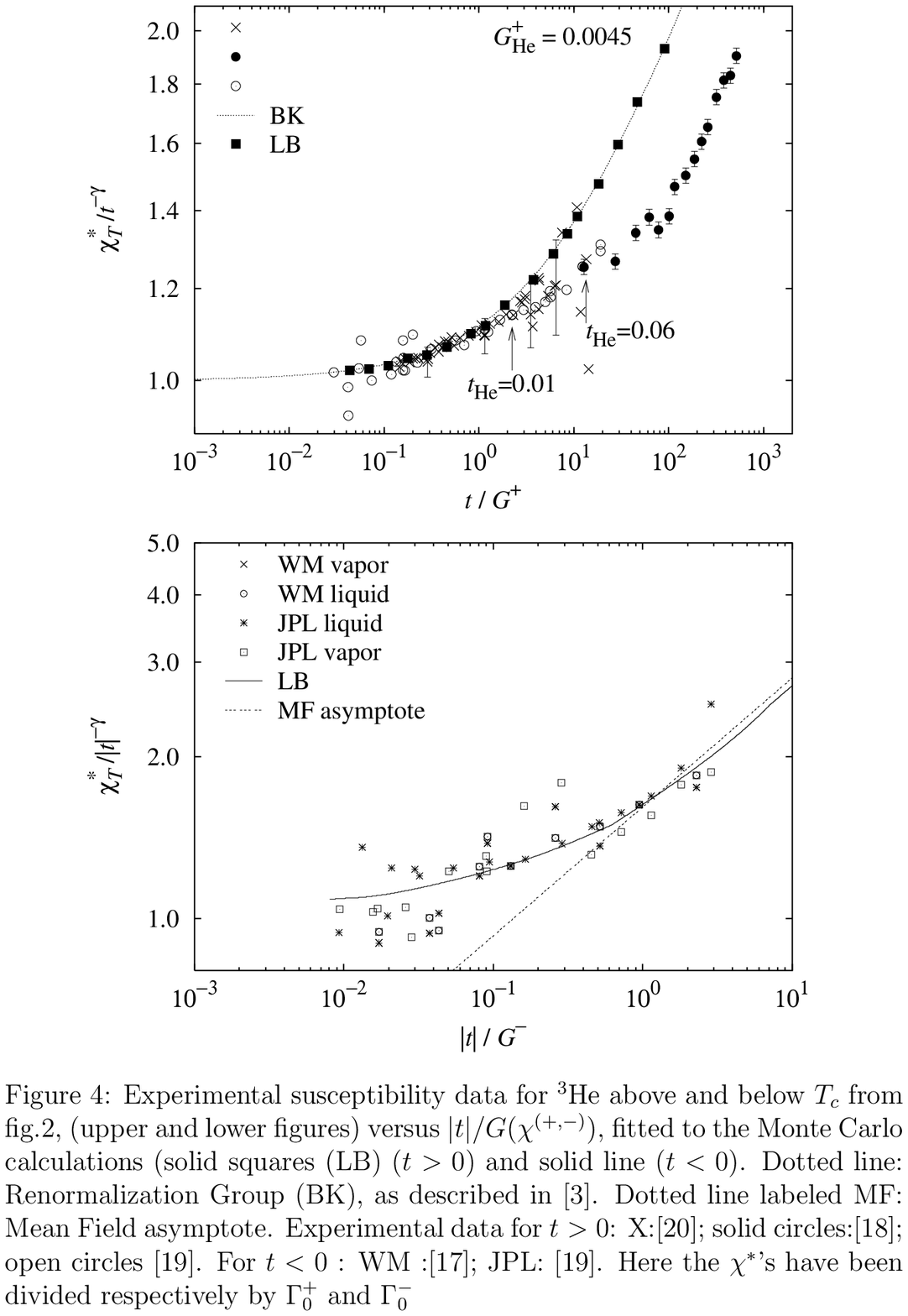,width=7in}}}
\label{fig:fig4}
\end{figure}

\section{Discussion} In Table 1, the values of the first correction term
amplitudes, obtained from a fit of the experimental data to Eqs.1 and 2
are obtained and their ratios are compared with predictions
\cite{Bagnuls:B:M:N:87}. The large (systematic) error bars reflect the
 data scatter and fit quality.  A fair consistency within the large
uncertainties is obtained, indicating that the result from the data
analysis appears consistent with the universality prediction based on the
$\Phi^4$ model.

In Tables 1 and 2,  the Ginzburg numbers and relevant ratios are listed,
and  the results are now briefly discussed. Here again the error bars
are  guesses based on the fitting uncertainties, since a satisfactory
error calculation could not be done. In Table 2, the G's are those
obtained by a fit of the data to the curve obtained from Monte Carlo
calculations. Obviously one of the great merits of MC calculations is to
give a much wider range of $|t|/G$ where the data  can be fit to
predictions than can a 2-term expansion such as Eqs 1 and 2. At the same
time, it is instructive to compare the resulting G's obtained from both
methods.

\begin{figure}[h]
\center{\parbox{7in}{\psfig{file=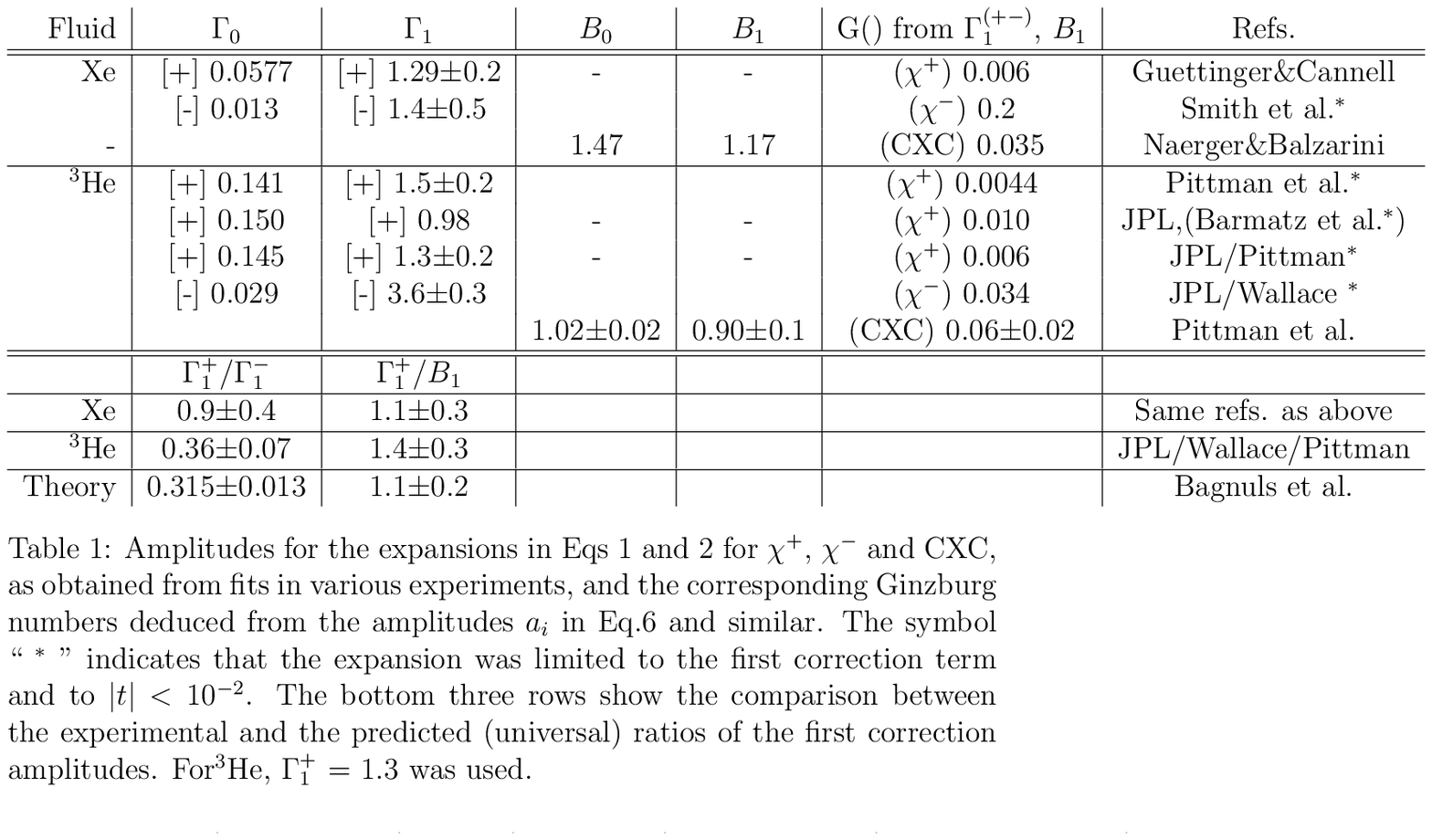,width=7in}}}
\label{fig:tab2}
\end{figure}

\begin{figure}[h]
\center{\parbox{7in}{\psfig{file=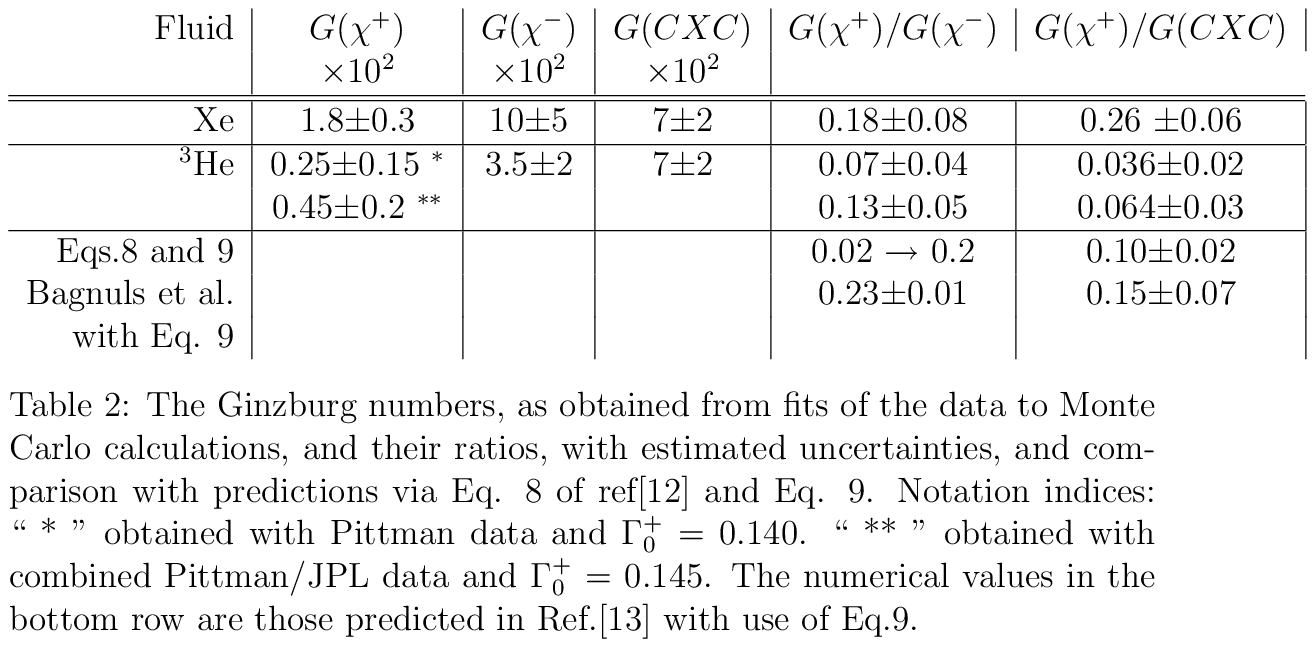,width=7in}}}
\label{fig:tab1}
\end{figure}

 This is done by comparing the numbers in the last column of Table 1 with
those on the first three columns in Table 2. On the whole, there is
acceptable  consistency between the determination from both methods, with
the exception for Xe. This  might  be caused by the different choices of
$\Gamma_0$ for the the $\chi^+$ data, as mentioned before.

We note that the Ginzburg numbers have uncertainties that reflect the
degree of difficulty in fitting the data to the curve obtained from Monte
Carlo calculations.  Yet, even with  this caveat, certain
 tentative  conclusions can be reached. First, the order of magnitude of
the
$G_i$ is as expected\cite{Anisimov:S:00}, namely $O(10^{-2})$. Second,
the ratios
$G(\chi^+)/G(\chi^-)$ are roughly the same for Xe and $^3$He, within the
stated uncertainty. From \cite{Aharoni:A:80} it is not clear whether there
should be universality for this ratio or for
$G(\chi^+)/G(CXC)$, where the experimental values for both fluids are
different. But as noted before, universality for the ratios of the first
correction term amplitudes is predicted from the $\Phi^4$
model\cite{Bagnuls:B:M:N:87}, and therefore the ratios of the
corresponding Ginzburg numbers, obtained via Eq.9, are universal too.

We now compare the ratios with those expected, based on
ref\cite{Aharoni:A:80}. For
$^3$He, the measured effective exponents for $\chi$  were
$\gamma_{eff}^+$ = 1.19
\cite{Wallace:M:70,Chase:Z:76,Pittman:D:M:79} and $\gamma_{eff}^-$ = 1.08
\cite{Wallace:M:70,Chase:Z:76} over the range
$5\times10^{-4}<|t|< 5\times 10^{-2}$ . When these $\chi$ data were
published, this result was  very surprising, because it was expected that
the exponents should be the same both above and below $T_c$. However in
the light of the Monte Carlo calculations that  show the crossover to be
quite different on both sides of $T_c$ (See Fig.1), this discrepancy in
the values of
$\gamma_{eff}$ can be understood. Interestingly the susceptibility data
for Xe\cite{Smith:G:B:71} do not show this difference, and both effective
exponents are listed\cite{Smith:G:B:71} as $\gamma_{eff}^{(+,-)}$ = 1.21
over the range $2 \times 10^{-4} <|t|< 8\times 10^{-3}$. For the
coexistence curve, the effective exponent has been reported to be
$\beta_{eff}$ = 0.360 for $^3$He
\cite{Wallace:M:70,Chase:Z:76,Pittman:D:M:79} and 0.355 for
Xe\cite{Cornfeld:C:72}.

The predicted ratios  $G(\chi^+)/ G(CXC)$ and $G(\chi^+)/G(\chi^-)$ from
Eqs.8 and 9, and from Bagnuls \underline{et}
\underline{al.}\cite{Bagnuls:B:M:N:87} via Eq.9 are listed in Table II.
Starting with the results from ref.\cite{Aharoni:A:80},
$G(\chi^+)/ G(CXC)
\approx$ 0.10 is determined from the effective exponents and lies in
between the values listed for Xe and
$^3$He. This prediction, which is
consistent with the value obtained by Bagnuls \underline{et}
\underline{al.}, is then good to within say $\pm$ 20\%. There
is agreement within the combined uncertainties for
$^3$He , but not so for Xe. The prediction of the ratio
$G(\chi^+)/G(\chi^-)$ from ref.\cite{Aharoni:A:80} with Eq.8,  is
uncertain : if $\gamma^+_{eff} = \gamma^-_{eff}$ is taken, as appears to
be the case for the Xe data,  the ratio is 0.2. However when  the values
for
$\gamma^+_{eff}$  and  $\gamma^-_{eff}$ for
$^3$He are used, as listed above, the ratio becomes 0.02 ! The first value
is consistent with the predictions by Bagnuls \underline{et}
\underline{al.}. Overall the experimental data  analysis in terms of the
correction term amplitudes
$a_i$ and the $G_i$'s is still in a  preliminary state and further
progress is needed.

\section{Conclusion} A status report has been presented of the program
describing  the crossover from asymptotic to mean-field behavior in
different properties for two simple fluids. So far, our understanding is
incomplete, since the accuracy of several sets of experimental data needs
substantial improvement. By contrast, MC calculations\cite{Luijten:B:98}
are making precise predictions of the crossover for the susceptibility
$\chi^+$ and
$\chi^-$ as well as the coexistence curve in terms of Ginzburg numbers.
Also there are quantitative predictions of the ratios of the correction
term amplitudes\cite{Bagnuls:B:M:N:87}.

In spite of the uncertainty in experimental data, some conclusions can be
reached. The Ginzburg numbers for $\chi$ and CXC in $^3$He and Xe   and
their ratios have been obtained from a data analysis. The latter were
compared with  predictions and discussed. Also from the ratios of the
first correction term amplitudes $B_1$,
$\Gamma_1(\chi^+)$ and
$\Gamma_1(\chi^-$), there appears confirming evidence of their predicted
universality
 within the large uncertainties. Further progress can be expected when
better experimental data of  the susceptibility of Xe and $^3$He below
$T_c$, and over a larger temperature range have been obtained.

\section{Acknowledgments} This research was supported by NASA grant NAG
3-1838. The greatest debt of gratitude goes to E. Luijten for the very
stimulating and informative interactions with him and for his generous
effort in preparing several plots. Furthermore he made a detailed
criticism of this paper.  I am also  indebted to A. B. Kogan for his
help with the plots in Fig.2, with data fitting and for his technical
help with the formatting,  to G.O Zimmerman for supplying an original
figure of the Chase and Zimmerman
$\chi$ data, to M. Barmatz and F. Zhong for permission to use the
unpublished $\chi$ data (labeled Zhong and JPL in the figures) obtained in
the JPL MISTE project, and to M. Giglio for supplying tabulations of
$\chi^-$ in Xe. I am very grateful to J.M.H. Levelt Sengers for
correspondence on the V.d.W. model and to F. Zhong for comments on this
report and for very useful suggestions. Finally I am indebted to C.
Bervillier for correcting some references, and to him and to E.
Vicari and A. Pelissetto for bringing those in
\cite{Bervillier} to my attention.

\section{Appendix}
\subsection{Extension of the rectilinear diameter above $T_c$ ?} In the
course of the data analysis for obtaining the $\chi^+$ of Xe from
ref.\cite{Michels:W:L:54}, the location of the maximum for $\chi^+$ with
respect to the critical isochore was determined. This line of points
might be thought to extend the trajectory of the rectilinear diameter as
T increases and passes the critical point. Over the whole range of the
data ($t<$ 0.47) it  was found to have a slope $\Delta \rho/t$ =  - 0.049,
 to be compared with the slope of  - 0.725 for the rectilinear
diameter\cite{Cornfeld:C:72}. In the Ising model, the slope is
zero for both lines. Similarly the Van der Waals model predicts the
rectilinear diameter slope as -2/5
\cite{Sengers:LS:78}, and from an expansion above
$T_c$ the slope for the maximum of $\chi^+$ to be zero \cite{Levelt:70}.
Hence there is a slope discontinuity in the same direction as observed in
Xe.
 For $^3$He, experiments give a rectilinear diameter slope of  +0.022
\cite{Pittman:D:M:79}. Above $T_c$  from an inspection of the the maximum
location in $\chi^+$ in the data analysis of various experiments
\cite{Meyer:99}, the slope is found to be zero within experimental error
over the range $t< 2\times 10^{-1}$. Beyond this range, the
$\chi^+$ versus $\rho$ curve along an isotherm is no longer symmetric with
respect to $\rho_c$ and as $t$ increases, the maximum of $\chi^+$ shifts
to larger densities.

\end{document}